\documentclass[twocolumn]{aastex701} 

\usepackage{natbib,graphicx,amsmath,amsthm,ulem,color,verbatim,enumerate}

\newcommand{\be}{\begin{eqnarray}}
\newcommand{\ee}{\end{eqnarray}}
\newcommand{\beqn}{\begin{eqnarray}}
\newcommand{\eeqn}{\end{eqnarray}}
\newcommand{\bi}{\begin{itemize}}
\newcommand{\ei}{\end{itemize}}
\newcommand{\ben}{\begin{enumerate}}
\newcommand{\een}{\end{enumerate}}

\newcommand{\dd}{{\partial}}

\newcommand{\E}{Z_{\rm free}}

\newcommand{\bld}[1]{\mbox{\boldmath$#1$\unboldmath}}

\def\pomega{\varpi}

\def\s{\, \rm s}
\def\K{\, \rm K}

\def\s{\, \rm s}

\def\m{\, \rm m}

\def\g{\rm g}
\def\pomega{\varpi}

\renewcommand{\E}{{\cal E}}
\newcommand{\A}{{\cal A}}
\newcommand{\B}{{\cal B}}
\newcommand{\I}{{\cal I}}
\renewcommand{\K}{{\cal K }}

\newcommand{\black}{\color{black}}

\begin{document}

\title{Eccentric Disks With  Self-Gravity}

\author[orcid=0000-0003-4450-0528,gname=Yoram,sname=Lithwick]{Yoram Lithwick}
\affiliation{Department of Physics \& Astronomy, Northwestern University, Evanston, IL 60202, USA}
\affiliation{Center for Interdisciplinary Exploration \& Research in Astrophysics (CIERA), Evanston, IL 60202, USA}
\email{y-lithwick@northwestern.edu}

\author[orcid=0000-0002-6246-2310,gname=Eugene,sname=Chiang]{Eugene Chiang}
\affiliation{Department of Astronomy, Theoretical Astrophysics Center, and Center for Integrative Planetary Science, University of California Berkeley, Berkeley, CA 94720-3411, USA}
\affiliation{Department of Earth and Planetary Science, University of California Berkeley, Berkeley, CA 94720-4767, USA}
\email{echiang@astro.berkeley.edu}

\author{Leon Mikulinsky}
\affiliation{Department of Mathematics, 970 Evans Hall, University of California Berkeley, Berkeley, CA 94720-3840}
\email{mikulinsky@berkeley.edu}

\author{Zhenbang Yu}
\affiliation{Department of Physics, 366 Physics North MC 3700, University of California Berkeley, Berkeley, CA 94720-7300}
\email{roger_yu@berkeley.edu}

\begin{abstract}  

Can a  disk orbiting a central body
  be eccentric, when the disk   feels its own  self-gravity and is pressureless? 
Contradictory answers appear in the literature.
  We show that to linear order in eccentricity such a
  disk 
  {\it can} be eccentric,  but only if it has a   sharply truncated  edge: 
 the surface density $\Sigma$ 
must vanish at the edge, and
 the  $\Sigma$ profile must  be sufficiently steep at the point where it vanishes. 
If  either requirement is   violated, an eccentric disturbance  leaks out of the bulk of the disk into the low density edge region, 
and cannot return. 
An edge where $\Sigma$  
 asymptotes to zero but  never vanishes, 
 as is often
assumed for astrophysical disks, is insufficiently sharp.
Similar results were
 shown by Hunter \& Toomre (1969) for galactic 
warps.
We demonstrate these results   in three ways:  by   solving the eigenvalue equation for the eccentricity profile; 
by solving the initial value problem; 
and by 
 analyzing a new and simple dispersion relation
 that is valid for any wavenumber, unlike WKB.
 As a byproduct, we  show that 
 softening 
 the self-gravitational potential 
 is not needed to model a flat disk, 
 and we
 develop  a softening-free algorithm to model the disk's Laplace-Lagrange-like equations. 
The algorithm is easy to implement and
is more accurate than softening-based methods at a given resolution
 by many orders of magnitude.

\end{abstract} 

\section{Introduction}

We consider astrophysical disks that orbit a  central body, and  that are  sufficiently
massive that self-gravity is important. 
Examples include planetary rings, protoplanetary 
disks, 
and disks of stars in galactic nuclei.
Many such disks are observed to be eccentric.
Understanding 
why and how such disks are eccentric 
should teach us about
their internal physics 
(e.g.~gravity, pressure, viscosity), and about  external perturbers, 
such as planets in protoplanetary disks, 
moons in planetary rings, and the galactic tidal potential of galactic nuclei.

Test particles orbit a central point  mass
along non-precessing Keplerian ellipses. 
A disk of test particles on nested ellipses  
 remains  eccentric forever.   
 But for a disk of massive particles the ellipses
interact, forcing one another to precess. Differential precession would rapidly  wipe out an
  eccentricity profile.  
Nonetheless, it is plausible that particular arrangements of eccentricity and mass can be made to precess uniformly, leading to a long-lived eccentricity. 
\cite{1979AJ.....84.1638G} worked out possible such profiles for the eccentric and radially narrow Uranian rings. Our goal here is to expand the class of such rigidly precessing solutions and understand more generally their properties.

  \cite{1999ASPC..160..307S}  solved the problem of an eccentric  and radially extended  disk
  of massive particles, with the ultimate
  goal of explaining 
  double nuclei of galaxies \citep{1995AJ....110..628T}.
They employed Laplace-Lagrange theory, in which  uniformly precessing
eccentricity patterns emerge as solutions of an eigenfunction problem  \citep{1999ssd..book.....M}.  
They successfully found solutions, 
as  may have been expected from Laplace-Lagrange theory.

\cite{2001AJ....121.1776T}  investigated self-gravitating eccentric disks by solving
the fluid equations rather than 
 using 
Laplace-Lagrange theory. 
 While he found that narrow rings can support eccentricity modes supported purely by self-gravity, he concluded that extended disks cannot, 
in contradiction to \cite{1999ASPC..160..307S}.
 We will resolve this disagreement as part of our study. 
 \cite{2001AJ....121.1776T} also explored the effect of softening gravity as a  model for pressure, and 
 developed
 a simple graphical technique for explaining the properties of the eigensolutions, 
 which we shall adapt below. 

Many later papers  studied
eccentric 
self-gravitating disks, 
sometimes 
including pressure or external perturbers such as planets
\citep[e.g.][]{2002A&A...388..615P,2002MNRAS.333..583T,2003ApJ...595..531H,
2015ApJ...798...71S,2016MNRAS.458.3221T,
2018ApJ...864...74D,
2019MNRAS.489.4176S,2019MNRAS.490.4353T}. 
\citet[][written with one of us]{2019ApJ...872..184L} 
found that  extended 
 disks with both pressure and self-gravity 
 do support 
  eccentric 
  modes. But as the strength of self-gravity is increased relative to pressure,
 the eccentric modes  
  become increasingly pathological, 
in the sense that their
 wavelengths  become 
  extremely small. 
 This 
accords with 
  \cite{2001AJ....121.1776T}'s finding  
   that 
   extended
 self-gravitating disks
  cannot be eccentric, 
  but contradicts 
  the 
  expectation from Laplace-Lagrange theory 
  \citep{1999ASPC..160..307S}.
 
In this paper we shall show that 
the key 
factor in whether a self-gravitating disk can be eccentric 
is the shape of the disk's surface density profile near its edge.  Other factors, such as whether the disk is extended or narrow, do not affect that conclusion.

\section{Master Equation}\label{sec:master}

We assume  throughout that the disk is two-dimensional, that the only operative
forces are self-gravity and gravity from the 
central mass, 
and that eccentricities  are governed
by the linear secular equations.  We  discuss additional effects in Section \ref{sec:discuss}.

The linear secular equation of motion for an orbiting body  that is perturbed by another is, from
\cite{1999ssd..book.....M}
\be
\dot{z} 
&=& i\left(G\over M_*\right)^{1/2}{m'\over 4 a a'^{1/2}}
\left(\alpha^{3\over 2}b_{3\over 2}^{(1)}(\alpha)z-\alpha^{3\over 2}b_{3\over 2}^{(2)}(\alpha)z'   \right)
\label{eq:mdeom} 
\ee
where $z$ is the perturbed body's  complex eccentricity, 
\be
z&=& ee^{i\pomega}  \ ;
\ee
 $a$ is the perturbed body's semimajor axis; 
 $z'$ and $a'$ are
  the corresponding quantities for the perturber; $m'$ is the perturber's mass; $M_*$ is the central mass;   
the $b_s^{(m)}$'s are Laplace coefficients; and $\alpha$ is the ratio of semimajor axes.
When written in the above form, one need not specify whether $\alpha=a/a'$ or $a'/a$ 
because the product $\alpha^sb_s^{(j)}(\alpha)$ is unchanged when $\alpha\rightarrow 1/\alpha$. 
Note that our primed variables
 refer to the perturber, in contrast to
 \citeauthor{1999ssd..book.....M}'s   primed variables  which to refer to the exterior body.

In what follows we
  convert the equation of motion from discrete bodies to that of a continuous disk, and then discretize the
continuum equations. That seemingly circular route will lead us to a more accurate numerical 
treatment 
 than simply adopting Equation (\ref{eq:mdeom}).
For the continuum limit of Equation (\ref{eq:mdeom}), we consider $z$ to be a function of $a$, and
$z'$ and $m'$  to be functions of $a'$; 
replace
$m'\rightarrow 2\pi \Sigma(a') a'da'$, where $\Sigma(a)$ is the surface density profile; 
 define
  the scaled surface density
\be
\sigma\equiv \pi  \left({Ga^3\over M_*}\right)^{1/2}   \Sigma =  {\pi G\over \Omega } \Sigma \ ,
\label{eq:sig}
\ee
which has dimensions of speed, and use that to replace $\Sigma(a')$ with $\sigma(a')$;
change the independent
variable from $a$ to
\be
u=\ln a ,
\ee
and similarly from $a'$ to $u'=\ln a'$;
 and integrate over $u'$.
The resulting continuum equation of motion is
\be
\dot{z}(u) &=&
i{1\over a(u)} \int \left(K_1(u'-u)z(u)-K_2(u'-u)z(u')  \right)\sigma(u') du' 
\nonumber
\ee
which we call the master equation.
The two kernels are defined by
\be
K_m(v)&=&{1\over \pi}\int_0^\pi {\cos m\phi \, d\phi\over (2\cosh v - 2\cos\phi)^{3/2}}  
\label{eq:km}
\ee
and are related to the  Laplace coefficients  via
$K_m(v)= (1/ 2) \alpha^{3\over 2}b_{3\over 2}^{(m)}(\alpha)\vert_{\alpha=e^v}$.
Note 
 that the kernels   depend only on the difference $u'-u$
\citep{1971ApJ...166..275K}, which
 will prove
 convenient. 
To cut down on  brackets we adopt a shorthand  for the master equation,
\be
 \dot{z}&=&i{1\over a} \int \left(K_1z-K_2z'  \right)  \sigma' du'  \label{eq:eom} \ ,
\ee
where throughout this paper 
kernels without explicit arguments   represent $K_m\rightarrow K_m(u'-u)$.

\begin{figure*}[t]
\centering 
\includegraphics[width=.45\textwidth]{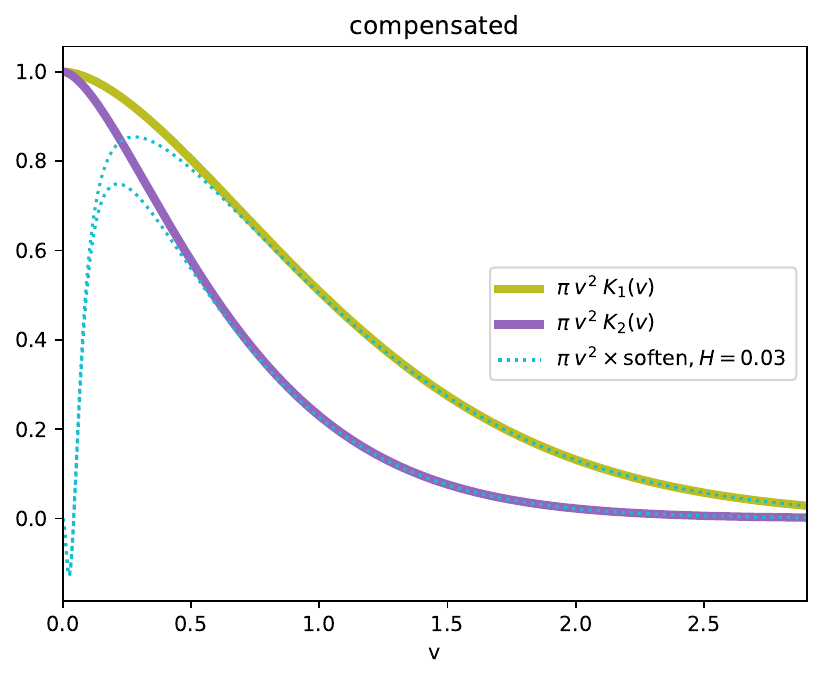}
   \raisebox{0.15cm}{ \includegraphics[width=.45\textwidth]{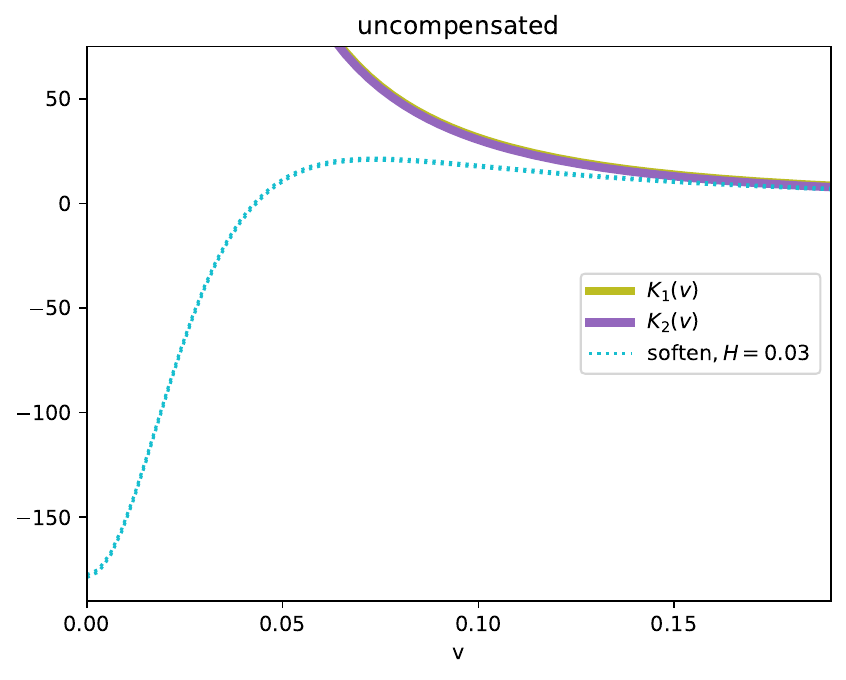}}
\caption{{\bf Kernels:} The kernels are defined in Equation (\ref{eq:km}). 
They
  have a strong  
$1/v^2$ divergence
at  $|v|\ll 1$ (Equation (\ref{eq:kdiv})), and hence
 have been multiplied in the left panel by $\pi v^2$ to compensate. 
The blue dotted curves in the left panel are softened kernels \citep{2003ApJ...595..531H} with the same compensation by $\pi v^2$. The right panel shows the uncompensated kernels,
with axes adjusted.
The softened kernels have a rapid
zero-crossing  at
$v\lesssim H$, and so
 numerical integrations that incorporate these softened kernels  require
extreme  resolution.
The reason for the rapid zero crossing is
that the softened kernels must come from taking
derivatives of the softened potential, rather
than from directly softening $K_1$ and $K_2$
\citep{2001AJ....121.1776T,2019MNRAS.489.4176S}.
 }
\label{fig:kernels}
\end{figure*}
The two  kernels are plotted in  Figure \ref{fig:kernels}.
  They are symmetric, $K_m(v)=K_m(-v)$.
At  $|v|\ll 1$ they diverge strongly,
\be
K_1(v)&\approx&K_2(v)\approx  {1\over \pi v^2} +O(\ln |v|) \ . \label{eq:kdiv}
\ee
where the $O(\ln |v|)$ term is different for the two $K$'s.
The kernels may be expressed in terms of associated Legendre functions,
 elliptic integrals, or hypergeometric functions \citep{1999ApJ...527...86C}. 

The master equation  conserves the angular momentum deficit
 \be
 {\rm AMD}&\equiv&\int |z|^2   \sigma a du \ ,  \label{eq:amd} 
 \ee
\citep[e.g.,][]{2016MNRAS.458.3221T}, 
which follows from the kernels' symmetry in $v$.

Many papers have studied
equations of motion similar to the master equation,
  including 
\cite
{1999ASPC..160..307S,2001AJ....121.1776T,2002A&A...388..615P,2003ApJ...595..531H,2016MNRAS.458.3221T,2019MNRAS.489.4176S} and \cite{2019ApJ...872..184L}.

\cite{1999ASPC..160..307S} modelled their disk
as discrete bodies, using Equation (\ref{eq:mdeom}) directly.
We shall show that
 this approach, called Laplace-Lagrange theory, converges
 to the  continuum solution as the number of discretizing bodies is increased. But the convergence is slow.
 
  \cite{2001AJ....121.1776T} derived our master equation (his Equation (42)) in a
  different way, by linearizing the two-dimensional fluid equations.
   \cite{2001AJ....121.1776T}  states that Laplace-Lagrange 
  (L-L) 
   theory is incorrect
 for a continuous disk, because it assumes eccentricities are small compared
 to the relative distance between rings, and that 
  that  inapplicable assumption may be rectified
 by softening. But the fact that the two derivations
 produce identical continuum equations shows that Laplace-Lagrange theory
 does not rely on such an extreme assumption. 
 While L-L theory does fail for crossing orbits (e.g.~for a test particle whose orbit crosses those of disk particles), it yields correct behavior for non-crossing orbits in the continuum limit, and in particular the correct (positive) precession rate for an  eccentric disk.

\section{Discretization Without Softening}
To solve the master equation, we  discretize  in $u$.  Our discretization
will  differ slightly from 
 Equation (\ref{eq:mdeom}). 
But that slight difference leads to  faster convergence to the correct answer as the number
of discretization points increases. 
The derivation of the discretization is in the appendix. Here, we provide a summary.
Readers uninterested in numerics
 may skip to the next section without  loss.

The integral in Equation  (\ref{eq:eom})
\be
\I(u)&\equiv & \int\left(K_1z-K_2z'  \right) {\sigma' }du'  \label{eq:idefpaper}
\ee
  must be evaluated carefully. The kernels diverge sufficiently strongly at
small argument $(|u-u'|)$ that each of the two terms in Equation (\ref{eq:idefpaper})
is infinite. 
But the two infinities
cancel, 
 because as $u'$ approaches $u$, then both 
 $z'\approx z$ and
 $K_1\approx K_2$ (Figure \ref{fig:kernels}); or to be more precise,  $K_1-K_2\approx O(\ln|u-u'|)$, and the integral
 of that difference does not diverge (see below Equation  (\ref{eq:gjk})).

\begin{figure*}[t]
\centering
\includegraphics[width=.7\textwidth]{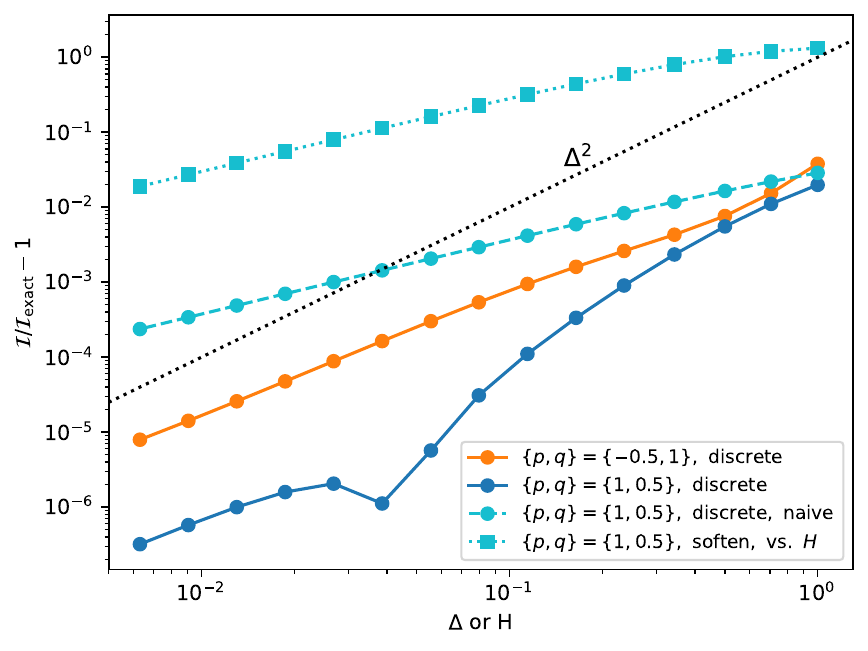}
\caption{{\bf Comparison of Approximations to  $\I$:}
The points connected by solid curves (orange and dark blue) show the error in the discretized
approximation (Equation (\ref{eq:idisc})), versus grid spacing $\Delta$.  
The error is better than 
$\Delta^2$. The light blue circles (labelled ``discrete, naive'') repeat the dark blue circles, 
 but with the factor of 1.5 in Equation (\ref{eq:fjk}) changed to unity. 
 The light blue squares  (labelled ``soften'') show the result with a softened kernel,
 integrated at infinite resolution ($\Delta\rightarrow 0$), 
 and plotted versus $H$ rather than 
 $\Delta$. 
This performs much worse than the dark blue circles.  For a fairer comparison, the light blue squares should be
 discretized at $\Delta\ne 0$, and
 plotted
 versus $\Delta$;
  that would  worsen the discrepancy with the dark blue circles. 
 }
\label{fig:Idisc}
\end{figure*}

An elegant way to avoid the infinities  is to soften the kernels,
which forces the $K_1$ and $K_2$ integrals to converge separately 
\citep{2001AJ....121.1776T,2002MNRAS.333..583T,2003ApJ...595..531H}. 
Example softened kernels are shown as   blue dotted curves
in Figure \ref{fig:kernels}.
 As the softening length is reduced, 
$\I$ converges to the correct answer, as shown by \cite{2019MNRAS.489.4176S}.  Softening has the added benefit that it mimics what
happens in a 3D disk.  But softening comes with an onerous computational cost.  
To achieve an error
of $\delta\I$ in an extended disk,
while adequately resolving the 
softening scale,
requires $\gtrsim 3(\I/\delta\I)^2$ gridpoints per decade
in $a$ \citep{2019MNRAS.489.4176S}.  For example, for a 1\% error more than 30,000
gridpoints per decade are required,\footnote{
When softened kernels are used, there are two contributions to the error $\delta\I$.  First, softening changes the kernel, 
and so  produces an error even if the integral over the softened kernel is performed
 at infinite resolution. In the example presented below (Figure \ref{fig:Idisc}),  the softening scale  must be $H<0.003$
in order to achieve a 1\% error at infinite resolution.
Second, there is an error due to discretizing the integral. 
Because of the rapid variation of the softened kernel
near $v\sim 0$ (Figure \ref{fig:kernels}), accurate discretization requires $\Delta\ll H$.  Extrapolating 
Figure 9 in \cite{2019MNRAS.489.4176S},  one sees that at least
30,000 gridpoints per decade are needed to achieve 10\% accuracy of the softened kernel at $H=0.003$; to  achieve
a 1\% error would require even  more gridpoints.
}
which is computationally challenging:
the eigensolution
of an $N\times N$ matrix on a laptop takes 
around 1 second$\times (N/3000)^3$, which
rapidly becomes problematic as $N$ is increased. The memory cost also explodes.

We avoid this difficulty by avoiding softening.\footnote{
\cite{1980Icar...41...76H} considers the perturbation
induced by a circular disk that has a power-law surface density
profile.  For that case, he is able to avoid softening by computing
the relevant integral analytically.
}
But we must then carefully account for the near-divergences in 
Equation (\ref{eq:idefpaper}). 
We do that
in the appendix.
We assume there that 
 $\sigma$ and $z$ are  intrinsically smooth functions of $u$;
  adopt
 a  grid in $u$ with uniform grid-spacing $\Delta$; and use the values
 of $\sigma$ and $z$ on that grid. 
The  result for the discretization of $\I$ is given by
Equation (\ref{eq:idisc}). Its error is $\sim \Delta^2$. Hence  to achieve a 1\% error in an extended disk it
suffices to take $\Delta\sim 0.1$, corresponding to $\sim 20$ gridpoints per decade in $a$.

We demonstrate the
accuracy of our discretization in Figure \ref{fig:Idisc}. We use the
power-law form of
 Sefilian and Rafikov,
 \be
\sigma &=& a^{-p+3/2} \\
z&=& a^{-q} \ ,
 \ee
and    evaluate $\I$ at $a=1$ ($u=0$).  We choose the bounds on the integral to be
$u'=-12,12$. 
The orange and dark blue points show the fractional error $\delta \I/\I$ at different grid spacings, where
the two colors are for different values of $p$ and $q$. One sees that the fractional error is better
than $\Delta^2$.  
The points labelled ``discrete, naive''
are evaluated in the same way as the corresponding discretized points, 
but with the factor of 1.5 in Equation (\ref{eq:fjk}) changed to unity, as in Equation (\ref{eq:naive}). That nearly
trivial
 change, which corresponds to
 dropping the single point with index $k=j$ (as in Laplace-Lagrange theory), makes the error  much worse; 
 but it still converges to the correct answer at small $\Delta$. 

We also  compare in the figure with the result from softening. 
A variety of different softenings have been introduced in the literature, as summarized 
in \citet{2019MNRAS.489.4176S}.
  See also \cite{2025arXiv250610812R}, who introduce a 
softening that mimics Gaussian-stratified
disks.
We adopt, somewhat arbitrarily, the  one introduced by \cite{2003ApJ...595..531H}. 
We replace the $K$'s in Equation (\ref{eq:idefpaper}) with their softened counterparts, 
which are provided in Table 1 of \cite{2019MNRAS.489.4176S}.  Specifically, we replace $K_1(v)\rightarrow 4\alpha^{1/2}\phi_{11}\vert_{\alpha = e^v}$, where the $\phi_{11}$ is listed in that table, and $K_2(v)\rightarrow -2\alpha^{1/2}\phi_{12}\vert_{\alpha = e^v}$. The $\phi$ functions depend both on $\alpha$ and
on a dimensionless softening length $H$, which is
roughly analogous to our grid spacing $\Delta$, in the sense that $H$ sets the lengthscale below
which  interparticle forces are modelled incorrectly
(Figure  \ref{fig:kernels}). Using those softened $K$'s, we perform 
 the integral in 
$\I$  numerically at infinite resolution, 
i.e., we ensure that the discretization scale $\Delta$ 
is sufficiently small that it does
not affect the accuracy of the integral.
The result is shown as the light blue squares in Figure \ref{fig:Idisc}, versus softening scale $H$.
Comparing with the dark blue circles shows that
 softening performs significantly worse. 
As mentioned above, the
performance of softening is even worse than suggested by the figure. 
In order to discretize the softened integral, 
one must use discretization $\Delta\ll H$.
Hence the light blue squares in Figure \ref{fig:Idisc} should be shifted
to the left by the factor $H/\Delta\gg 1$
when plotted versus $\Delta$
instead of $H$.

\section{Sharp Edge Versus Soft Edge}
\label{sec:twoedges}

\begin{figure*}
\begin{minipage}[b]{.9\textwidth}
    \centering
    \includegraphics[width=1.05\textwidth]{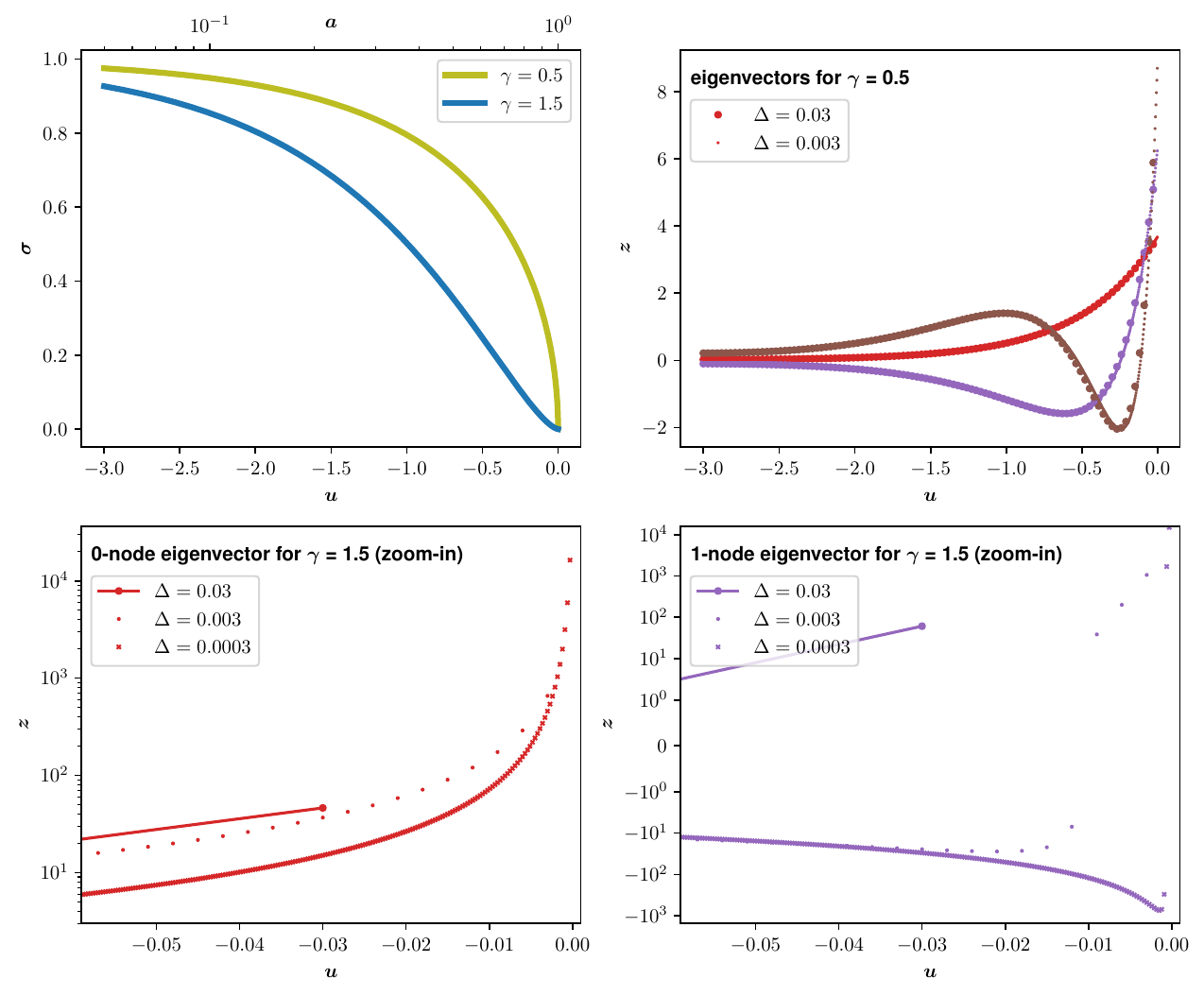}
 \end{minipage}
\caption{{\bf Disk Profiles and Eigenvectors}
}
\label{fig:evecs}
\end{figure*}

The main result of this paper is that 
disks with sharp edges can remain eccentric, whereas those
with soft edges cannot.  To highlight the dichotomy, in this section we solve the master equation for
two disk profiles that bracket the sharp vs. soft transition. 
For both profiles,  the scaled surface density
 (Equation (\ref{eq:sig})) follows the profile
\be
\sigma = 
(1 - a)^\gamma & \ \ \text{for } a < 1  \label{eq:profile} \ ,
\ee
with  an outer  edge at $a=1$. 
For the sharp profile,
  $\gamma=0.5$, and for the soft, $\gamma=1.5$. 
  These are
plotted in the top-left panel of Figure \ref{fig:evecs}.
  The two profiles look broadly similar, but will be shown to
    produce
   dramatically
different eccentricity behavior. 
One may imagine that Equation  (\ref{eq:profile}) extends to $a\rightarrow 0$. 
In truth, for the numerical experiments in the present section  
we set  the inner boundary to be at $a=0.05$ ($u=-3$), and  set   $\sigma=0$ inside
of that.  But we will verify in Section \ref{sec:other}  that the inner boundary plays a negligible role.

In the following, we first 
 solve
 the eigenvalue problem, and then evolve in time. We  turn to  the  theoretical explanation 
 in
  Section \ref{sec:theory}.

\subsection{Eigensolution}
We solve the master equation (Equation (\ref{eq:eom})) by setting
$z(u,t)=\hat{z}(u) e^{i\omega t}$.
That produces an eigenvalue
equation, with eigenvalue $\omega$, and eigenvector $\hat{z}$.
The discretized eigenvalue problem is derived in the appendix, and given
in matrix form in
Equation
(\ref{eq:eommatrix}). 
As shown there, the eigenvalues and
eigenvectors are  real-valued. The eigenvalues are  positive ($\omega>0$), which corresponds
to
prograde precession.

The top-right panel in Figure \ref{fig:evecs} shows the first three  eigenvectors for the 
sharp ($\gamma=0.5$) case,
where eigenvectors are ordered  by the number
of nodes, or equivalently by their eigenfrequencies, which increase with node number. 
The eigenvectors are seen to converge with resolution: as  the grid spacing is reduced
from $\Delta=0.03$ to 0.003, they hardly change.

By contrast, for the soft ($\gamma=1.5$) profile the eigenvectors
are pathological, as shown in the bottom panels of Figure \ref{fig:evecs}. 
As $\Delta$ is decreased, they become peaked ever closer
to the outer edge.
We normalize our eigenvectors
to have AMD=1 (Equation (\ref{eq:amdnorm})).  Hence the shift towards the edge seen in  Figure \ref{fig:evecs} 
reflects a shift of AMD.
The eigenvectors
  have not converged even for $\Delta=0.0003$. Based on the trend seen in the figure, 
   they presumably will never converge.

\begin{figure}[h]
\centering
    \includegraphics[width=.5\textwidth]{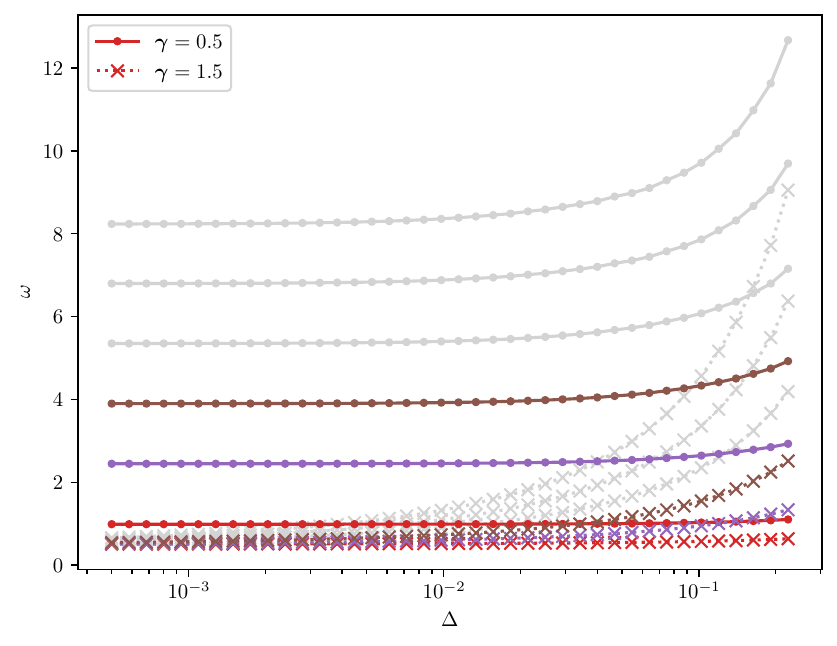}
\caption{{\bf Eigenvalues (mode frequencies)  vs. grid spacing:}
For each profile ($\gamma=0.5$ and 1.5), the first 6 eigenvalues are shown. 
For the $\gamma=0.5$ case, the eigenvalues are well separated at high resolution. 
But for the $\gamma=1.5$ case, they become degenerate at high resolution. 
 }
\label{fig:evals}
\end{figure}
A similar inference follows from the dependence of the eigenvalues on resolution (Figure \ref{fig:evals}).
For the $\gamma=0.5$ case, $\omega$ quickly converges to a fixed value as $\Delta$ decreases. 
But for the $\gamma=1.5$ case, all of the modes' frequencies approach each other.  At high resolution, 
the frequencies become degenerate.

\subsection{Time Evolution}

\begin{figure*}
    \centering
    \includegraphics[width=.465\textwidth, trim=0 0 80 0, clip]{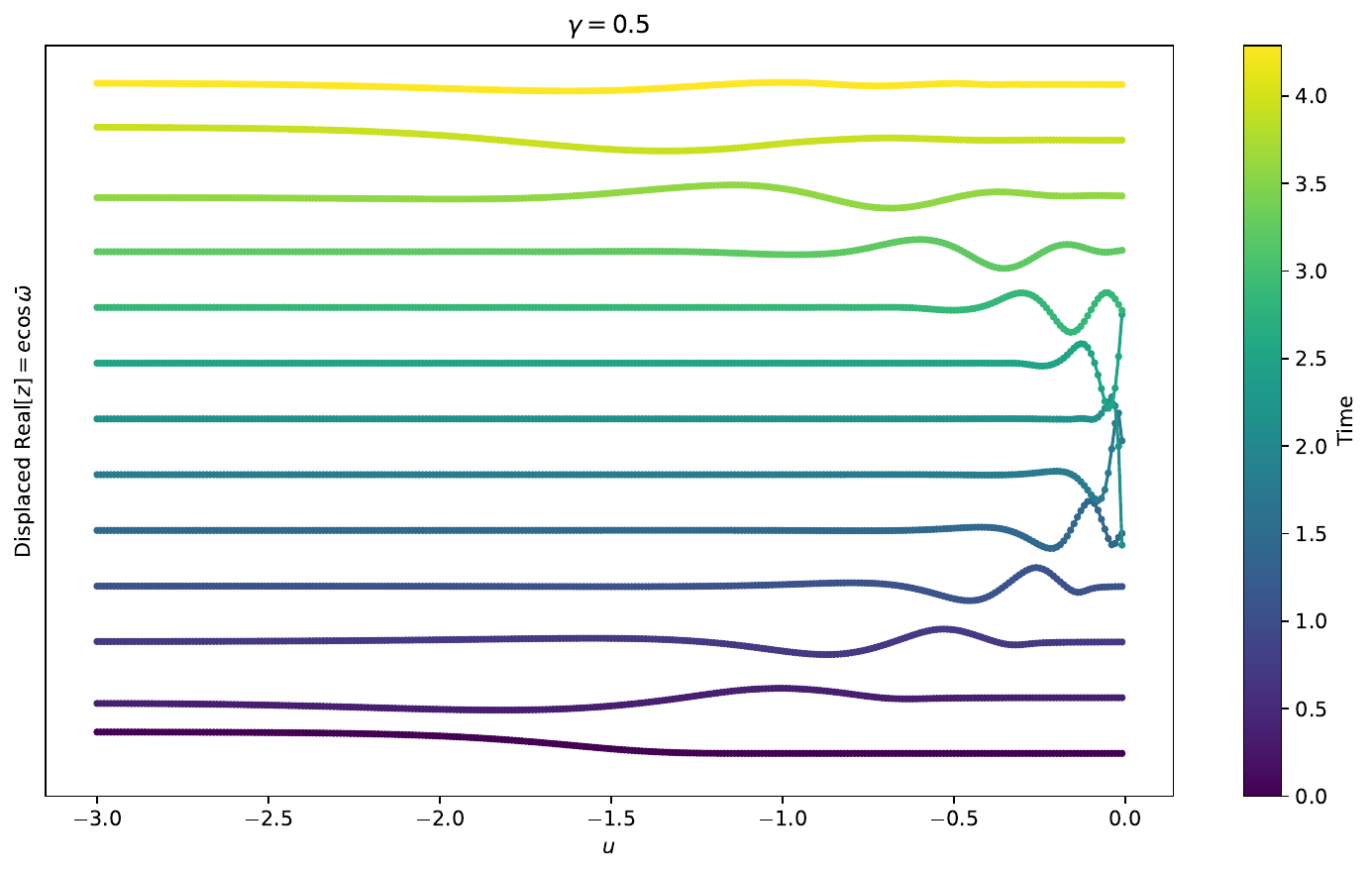}
  \includegraphics[width=.52\textwidth]{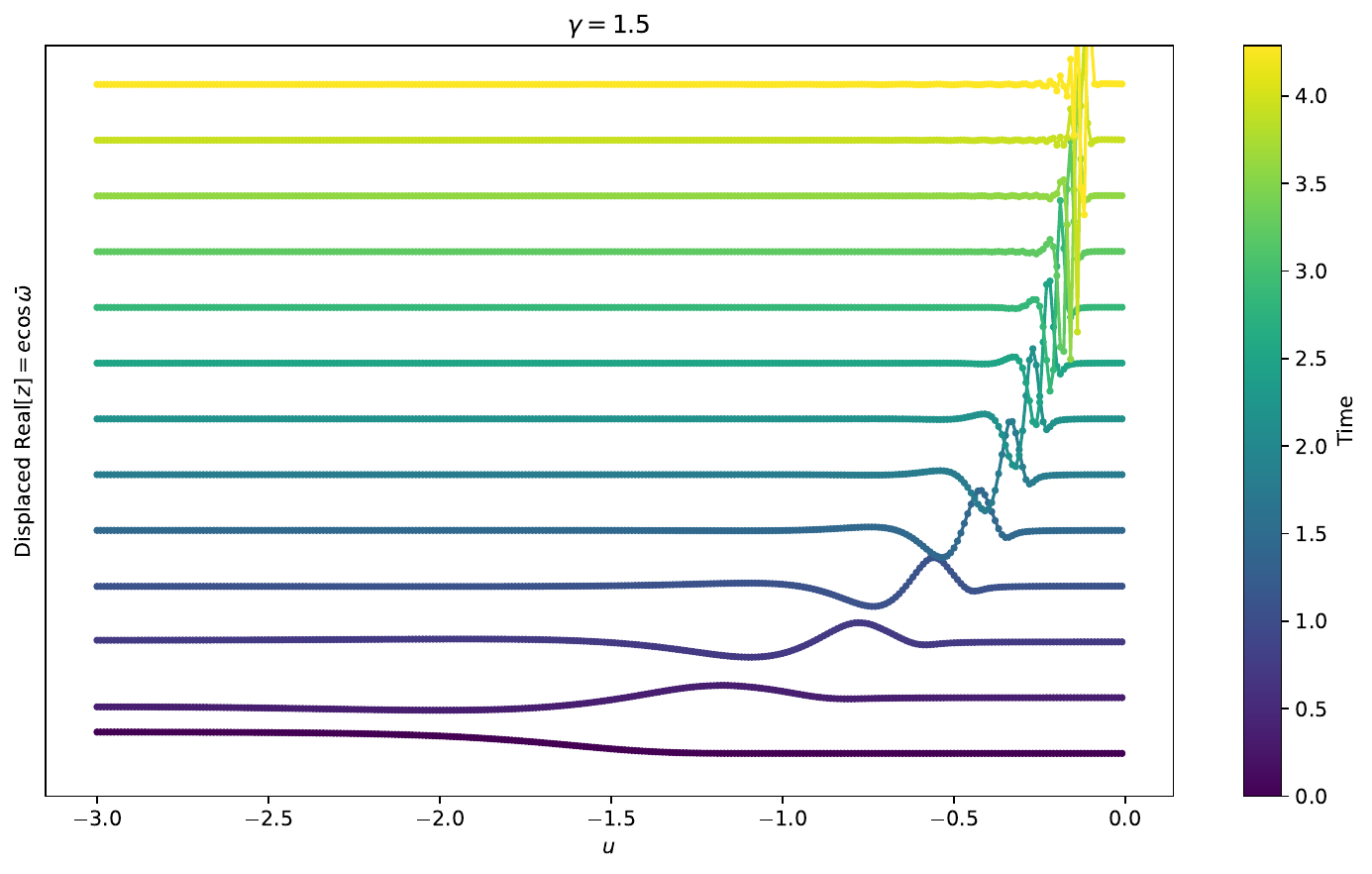}    
\caption{{\bf Time evolution:} For the $\gamma=0.5$ case, the disturbance initially travels outward, and then bounces off
of the outer edge. For  $\gamma=1.5$, it never reaches the outer boundary, and instead becomes
highly oscillatory. 
}
            \vspace{1.cm}
\label{fig:tevol}
\end{figure*}

We consider now the time-dependent problem. 
We initialize the eccentricity profile as
\be
z_{\rm init} =e_{\rm init}= e^{-(a/0.2)^4} \ ,
\ee
i.e., purely real (nodally aligned), with constant eccentricity
 at $a\lesssim 0.2$, and with a rapid transition to circularity for $a\gtrsim 0.2$. 
We then evolve $z$ forward in time by numerically integrating the discretized equation of motion
(Equation (\ref{eq:eomdisc})), with grid spacing $\Delta = 0.01$.  The results are shown in Figure
\ref{fig:tevol}, where we plot profiles of 
${\rm Real}[z]$ at successive times, displaced vertically. 
In the $\gamma=0.5$ disk, the initial eccentric  disturbance travels towards the outer edge, where it reflects
and then travels inwards. But in the $\gamma=1.5$ disk, the disturbance
stalls. It  produces
 increasingly rapid spatial variations  before it can reach the outer edge. At some point, the disturbance
 will become nonlinear, invalidating our assumption of linearity in $e$. To understand what happens next
 would require solving the nonlinear equations of motion, which we do not attempt in this paper.

\section{Theory }
\label{sec:theory}

The behavior observed in the previous section follows from the dispersion relation.
We derive the dispersion relation by adopting the ansatz
\be
z = {\rm const}\times {1\over \sigma(u)}e^{i(\omega t - k u)} \label{eq:logspir}
\ee
for $k$ assumed constant and real.
Aside from the $\sigma$ prefactor (to be discussed shortly), the 
spatial dependence is that of  a
logarithmic spiral: it has constant
wavelength
  in $u\equiv \ln a$ 
 \citep{2008gady.book.....B}. 
The temporal dependence corresponds
to the spiral precessing at angular frequency $\omega$. 
We shall find that $\omega>0$, which implies prograde precession.
Inserting this ansatz into the master equation  (Equation  (\ref{eq:eom})) gives
\be
{\omega\over\sigma(u)}e^{-iku} &=&
 {1\over a(u)}\int_{-\infty}^\infty \left( K_1{\sigma(u')\over \sigma(u)}e^{-iku}-K_2e^{-iku'} \right)du' \ ,
\nonumber
\ee
where we again suppress the dependence of the $K$'s  on $u'-u$ for clarity.
Multiplying through by $e^{iku}$ and replacing $K_2e^{ik(u-u')}\rightarrow K_2\cos(k(u'-u))$
due to the symmetry of $K_2$, 
 we arrive at
 the dispersion relation 
\be
 \omega(k,u)&=& {\sigma(u)\over a(u)}\K(k) + \nu(u)
 \label{eq:dr}
\ee
where
\be
\K(k)&\equiv& 2\int_0^\infty K_2(v)(1-\cos k v)dv
\label{eq:kap}
 \\
\nu(u)
&\equiv & {1\over a(u)}\int_{-\infty}^\infty \left(K_1{\sigma}(u') - K_2\sigma(u)\right) du'
\label{eq:nudef}
\ee
We have added a $K_2$ term  to Equation
(\ref{eq:kap}) and subtracted it from Equation (\ref{eq:nudef})  in order to
ensure that $\K$ and $\nu$ are each finite. 
We call $\nu(u)$ the  precession rate,
because 
it will be seen to be equal to $\omega$
at a mode's turning point. 
\cite{2001AJ....121.1776T}
defines the precession rate differently, as just the $K_1$ term in  Equation (\ref{eq:nudef}), or equivalently
as the coefficient of $iz$ in the master equation. That expression diverges 
if the softening is strictly zero.\footnote{Though it converges in the limit that the softening parameter approaches, but does not actually reach, zero.}

The above dispersion relation is very simple, in that there are no complex numbers, and its dependencies on $k$ and
$u$ are clearly separated.
It is an exact consequence of the master equation, subject only to the assumptions that $k$ be constant and real.\footnote{
When we apply the dispersion relation below, we will violate the assumption that
$k$ be constant, in effect assuming that it is  nearly constant.
} 
Unlike in WKB theory, one need not assume that $|k|$ be large. 
A different dispersion relation is derived in \cite{2019ApJ...872..184L}
 by expanding the equation of motion at large
$k$, 
to $O(1/k^2)$. But Equation (\ref{eq:dr}) is both simpler and valid at any $k$.
The simplicity relies on
two ``tricks'' that
 are standard in galactic dynamics
\citep{1971ApJ...166..275K}:
(i)  
the kernels are expressed as a function only of the difference of independent variables ($u'-u$), and (ii)    
$\sigma$ is factored out of the exponential ansatz  (Equation (\ref{eq:logspir})).
But  unlike what is often done in galactic dynamics \citep[e.g.,][]{1976PhDT........26Z,1979SJAM...36..407B,1998MNRAS.300...83E}, we do not  make any assumptions regarding the shape of the background
surface density profile $\sigma(u)$.
That  allows us to avoid the cumbersome
``Kalnajs kernel'' $K(\alpha,m=1)$, 
which is related to the Fourier transform
of our $K_m(v)$.

\begin{figure}[h]
\centering
            \vspace{.5cm}
    \includegraphics[width=.5\textwidth]{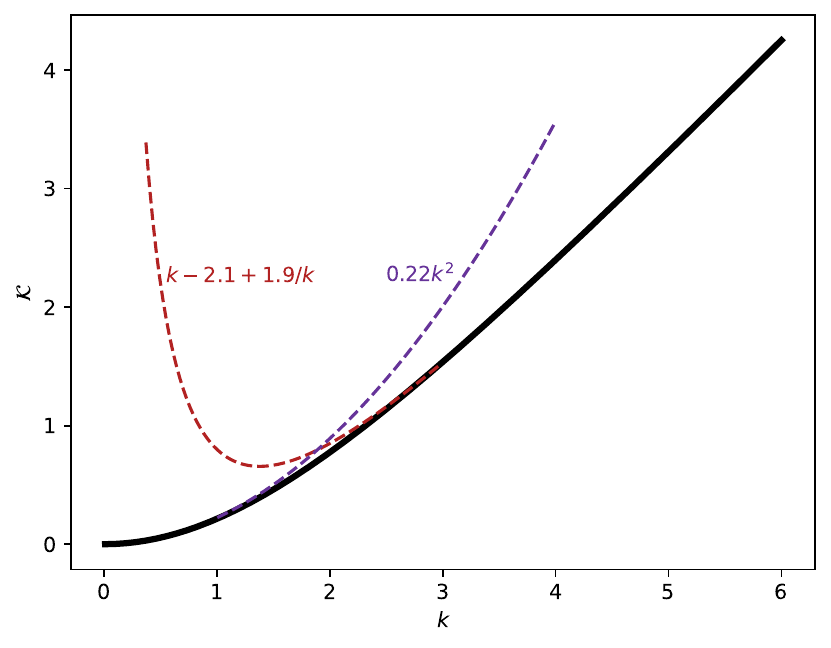}
\caption{{\bf Function $\K(k)$ in Dispersion Relation:}
The black curve shows $\K(k)$ (Equation (\ref{eq:kap})). 
We only show $k>0$ because it is symmetric, $\K(k)=\K(-k)$.
The dashed curves show its approximate form at high and low $k$.
The leading  term in the high $k$ expansion ($\K\approx |k|$) is derived
in the text, and the subdominant terms have been obtained by numerical integration. 
For  $|k|\ll 1$, one may replace $1-\cos k v\rightarrow (kv)^2/2$,
 which produces $\K\approx 0.223 k^2$. 
 }
\label{fig:kappa}
\end{figure}
The function $\K(k)$  in the dispersion relation is a universal function of 
$k$, independent of $\sigma$. It is plotted in Figure \ref{fig:kappa}.
Its behavior at  $|k|\gg 1$ is obtained
by replacing $K_2(v)\rightarrow 1/(\pi v^2)$
(Equation (\ref{eq:kdiv})), in which case $\K\rightarrow  |k|$. 
Hence at large $|k|$, 
\be
\omega &\approx& {\sigma\over a}|k|+\nu  \ , \ \ {\rm for\ }|k| \gg 1 \\
& =& {\pi G\Sigma\over \Omega a}|k| +\nu 
\ee
This is nearly the same as the Lin-Shu dispersion relation, after setting the sound speed to zero, the azimuthal wavenumber $m_\phi=1$, 
 the epicyclic frequency to
 $\Omega$, and expanding in $\omega/\Omega$. But our expression
for $\nu$ is different from Lin-Shu, and our 
$k$ provides
 the wavelength in $u$ rather than in $a$
 \citep[see discussion in][]{2019ApJ...872..184L}.

Before addressing $\nu$, we may already see why 
soft edges fail to trap modes. 
The group velocity  is
\be
v_{\rm group}={\dd\omega\over \dd k}&=&{\sigma\over a}{d\K\over dk}  \\
&\approx& \pm {\sigma\over a} \ , \ \ {\rm for\ }|k| \gg 1 
\label{eq:kg1}
\ee
which corresponds to the speed ($du/dt$) of a wavepacket; the reason for taking the $|k|\gg 1$ limit will be clarified at the end of Section \ref{sec:eld}. The travel time from some initial $u_{\rm init}$ to the edge is
\be
T&=&
\int_{u_{\rm init}}^0 {1\over v_{\rm group}}du  \\
&\approx& \int_{u_{\rm init}}^0 {a\over \sigma} {du} \label{eq:conv} \ .
\ee
Only for sufficiently sharp edges will this time be finite, indicating that the 
wavepacket can bounce back from the edge.
\cite{1969ApJ...155..747H} explain this result in a slightly different  way: they show that the number
of nodes is infinite when the  integral in Equation (\ref{eq:conv}) diverges. 
For our fiducial profile (Equation (\ref{eq:profile})), we  have
\be
  T&\approx& \int_{u_{\rm init}}^0 {du\over |u|^\gamma} 
 \label{eq:travel}
\ee
after setting
$\sigma \approx |u|^\gamma$ and $a\approx 1$, which hold near the edge. 
Hence the time to reach the edge is finite only when $\gamma< 1$, and only such profiles can support an eccentric mode.

\subsection{Energy-level diagrams}
\label{sec:eld}
The
   dispersion relation  controls  more than just  the behavior near the edge. It
   also 
   dictates
   where  eccentric modes exist, how they are trapped,   
   and their eigenfrequencies.  
   We may see that graphically  by adopting
 the technique of  energy-level  diagrams 
from
quantum mechanics
\citep{2016MNRAS.458.3221T,2019ApJ...872..184L}.
As a reminder of the technique, we first  write the  Schr\"odinger equation  
in a form 
that parallels Equation (\ref{eq:dr}): 
\be
 E&=&-{\hbar^2\over 2m}{1\over \psi}{d^2\psi\over du^2} + V(u)  \ ,
 \label{eq:schro}
\ee
where the symbols here have their usual quantum-mechanical interpretation.
One may solve this equation for $\psi(u)$  in a bound state,  and for the
energy $E$ of the bound state.  In a given state, $\psi(u)$ is confined by turning points (i.e., inflection points),
which occur where $V=E$.  In the bound region, where $V<E$,
the solution is oscillatory 
(the squared wavenumber $k^2\equiv -(1/\psi)d^2\psi/du^2>0$), while in the unbound region ($V>E$) it is evanescent. To make the correspondence with  eccentric modes, 
the independent variables are related via
\be
\psi \leftrightarrow \sigma z \ , \label{eq:si}
\ee
because Equation (\ref{eq:logspir}) implies
$k^2=-1/(\sigma z)d^2(\sigma z)/du^2$. 
And 
comparing Equation (\ref{eq:schro}) with 
 Equation (\ref{eq:dr}), we see the 
 further correspondences
 \be
E&\leftrightarrow& \omega \\
 V&\leftrightarrow& \nu  \\
 -{\hbar^2\over 2m}{1\over \psi}{d^2\psi\over du^2}
 &\leftrightarrow& {\sigma\over a}\K(k) \ .
 \ee
The last correspondence is
not perfect, both because of the $u$-dependent prefactor ($\sigma/a$), and
because $\K\propto k^2$ holds only at low $|k|$ (Figure \ref{fig:kappa}).  But that imperfection does not affect the most important
quantity extracted from an energy
level diagram: the location of the turning points, i.e., where $k=0$.
In particular, one may plot the ``potential''
$\nu(u)$, and consider what happens for different
``energies'' $\omega$. Eccentric modes will
be trapped by the turning points which occur where
$\omega=\nu$. The
 sign of $\omega-\nu$ dictates whether the ``wavefunction'' $\sigma z$ is evanescent or oscillatory.

\begin{figure}
\centering
            \vspace{.45cm}
    \includegraphics[width=.5\textwidth]{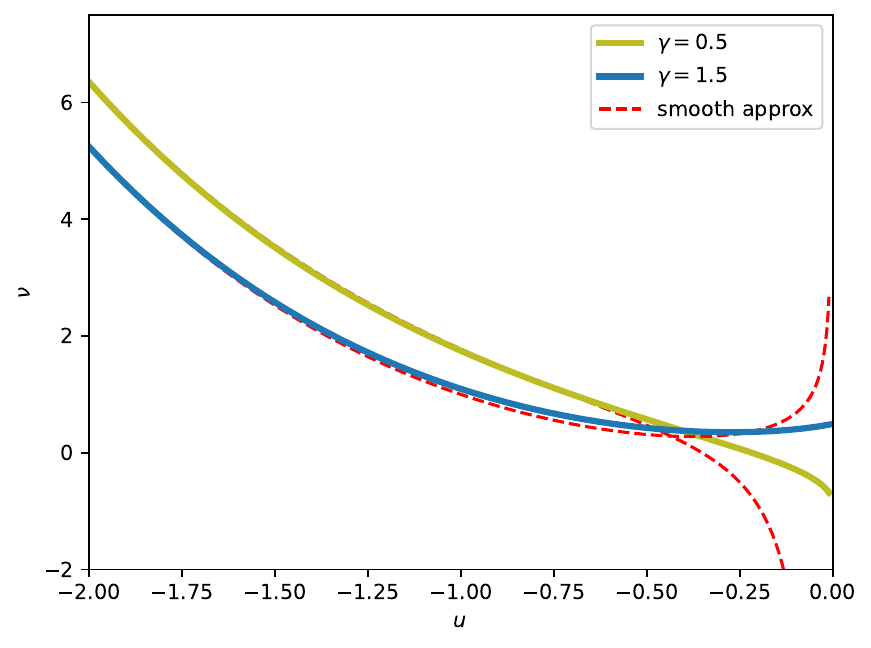}
\caption{{\bf Precession rates in our two fiducial profiles:}
The precession rate is defined in Equation (\ref{eq:nudef}).  
The ``smooth approximation'' is from 
Taylor-expanding $\sigma$ in Equation (\ref{eq:nudef}) to $O(u-u')^2$, which yields
$\nu \approx
{1\over a}\left( 0.96\sigma + 0.37 {d^2\sigma/ du^2} \right)$.
It is inaccurate near the edge, because $\sigma$ varies too rapidly there. 
 }
\label{fig:prec}
\end{figure}

We  now apply the technique to  the two fiducial profiles studied
in Section \ref{sec:twoedges}. 
In Figure \ref{fig:prec}, we show their effective potentials, $\nu(u)$. 
For the sharp edge (yellow), $\nu$ rises inwards, showing that a mode must
be confined to the outermost parts of the disk---it behaves as an edge mode. 
For the soft edge (blue), there is a shallow potential well slightly interior to the  edge.
We also show in the figure a simple approximation to $\nu$, which
 provides an adequate approximation everywhere except near the  edge. 

\begin{figure}
\centering
            \vspace{.5cm}
    \includegraphics[width=.5\textwidth]{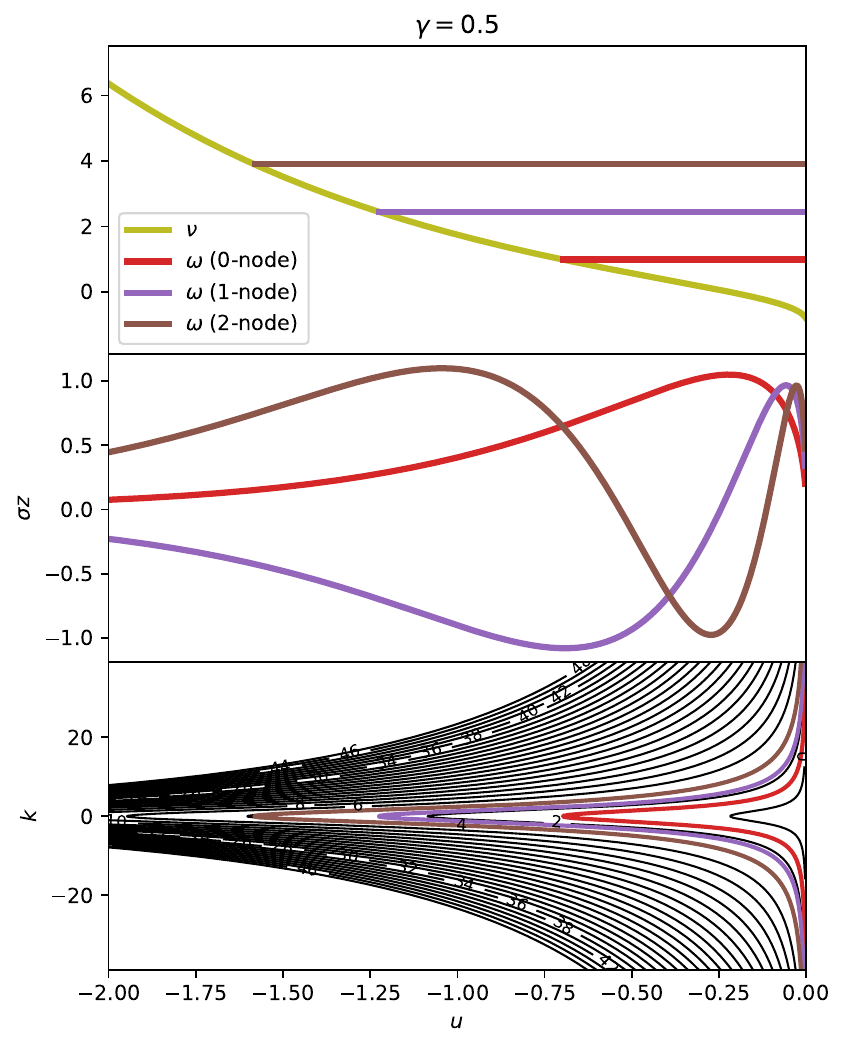}
\caption{{\bf Mode trapping in sharp-edged disk:}  The top panel shows the energy level diagram; 
the middle panel shows the eigenfunctions (scaled by $\sigma$); and the bottom panel shows
the  dispersion relation map.
 }
\label{fig:drm}
\end{figure}

Figure \ref{fig:drm} (top panel) shows the energy-level diagram for the sharp-edge case.
The function $\nu(u)$
there
 is repeated from Figure \ref{fig:prec},  and the horizontal lines show the frequencies of the first
three modes, shown previously in Figure \ref{fig:evals}.  
To illustrate that the turning points occur where $\nu=\omega$, we re-plot the eigenfunctions
(shown previously in Figure \ref{fig:evecs}), 
but now premultiplied by $\sigma$ because of Equation (\ref{eq:si}).
It may be seen that the crossing points in the top panel occur at the same place as the inflection
points in the middle one.

The bottom panel shows a ``dispersion relation map,'' which provides a complementary
view to the top panel 
\citep[e.g.][]{2008gady.book.....B,2001AJ....121.1776T, 2019ApJ...872..184L}.
Here the function
  $\omega(k,u)$ is displayed as a contour map. 
 The three colored curves have the same values of $\omega$
as the first three eigenfunctions that appear in the top panels.   Each of those curves is  the ``train track''
  along which the corresponding  mode is confined.

The  question of what happens near the outer edge is answered
by the shape of these curves near the right boundary. 
The shape is determined  by $\omega\approx |u|^\gamma |k|$, which follows
from the dispersion relation at large $|k|$. For a contour to represent a mode, 
one needs the area enclosed by the contour  to be $\oint k du=\pi, 3\pi, 5\pi, \cdots$ \citep{2019ApJ...872..184L}. 
Hence with $k\propto |u|^{-\gamma}=|u|^{-.5}$, the area within any of the contours is finite, leading
to well-separated frequencies.  The case $\gamma=1.5$ will  lead to the opposite conclusion.  
One may see also from this panel that our earlier assumption that $|k|\gg 1$ near the edge (Equation (\ref{eq:kg1})) is justified.

\begin{figure}
\centering
            \vspace{.5cm}
    \includegraphics[width=.5\textwidth]{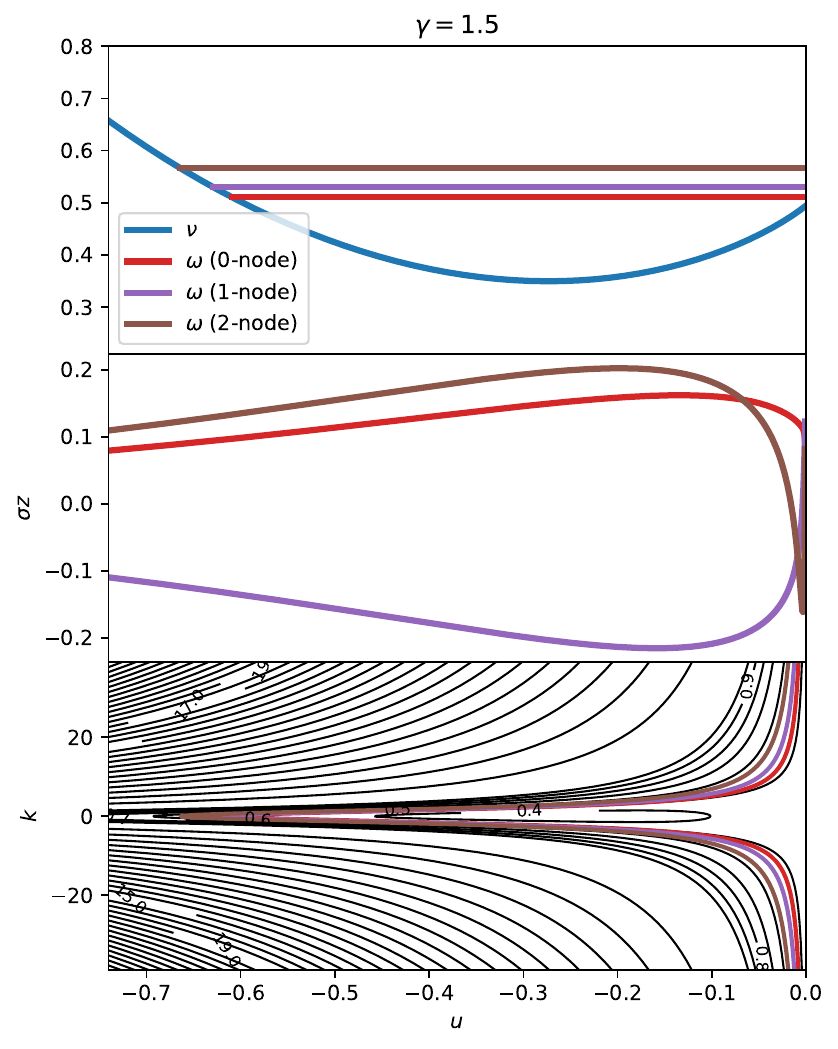}
\caption{{\bf Mode Trapping in Soft-Edged Disk.} 
 }
\label{fig:drmgamhi}
\end{figure}
Figure \ref{fig:drmgamhi} shows what happens for the $\gamma=1.5$ case.  
The eigenvalues and
eigenvectors shown here are calculated with $\Delta=0.001$. 
From the bottom panel, one sees that the colored
contours near the outer boundary are much steeper than for the $\gamma=0.5$ case.  As explained above, these
here have functional form $k\propto |u|^{-1.5}$, and so the area enclosed by those contours, $\oint k du$, does not converge.

\section{Other Disk Profiles}
\label{sec:other}

We examine the eccentricity solutions for two additional kinds of  density profiles. 
 For the first kind, we examine the role
of an inner edge.  We take $\sigma(u)$  to be the same as our fiducial profile with
a sharp edge ($\gamma=0.5$, Equation (\ref{eq:profile})), but now we  sharply truncate $\sigma$
to zero at  an inner edge.
\begin{figure}
\centering
            \vspace{.5cm}
    \includegraphics[width=.5\textwidth]{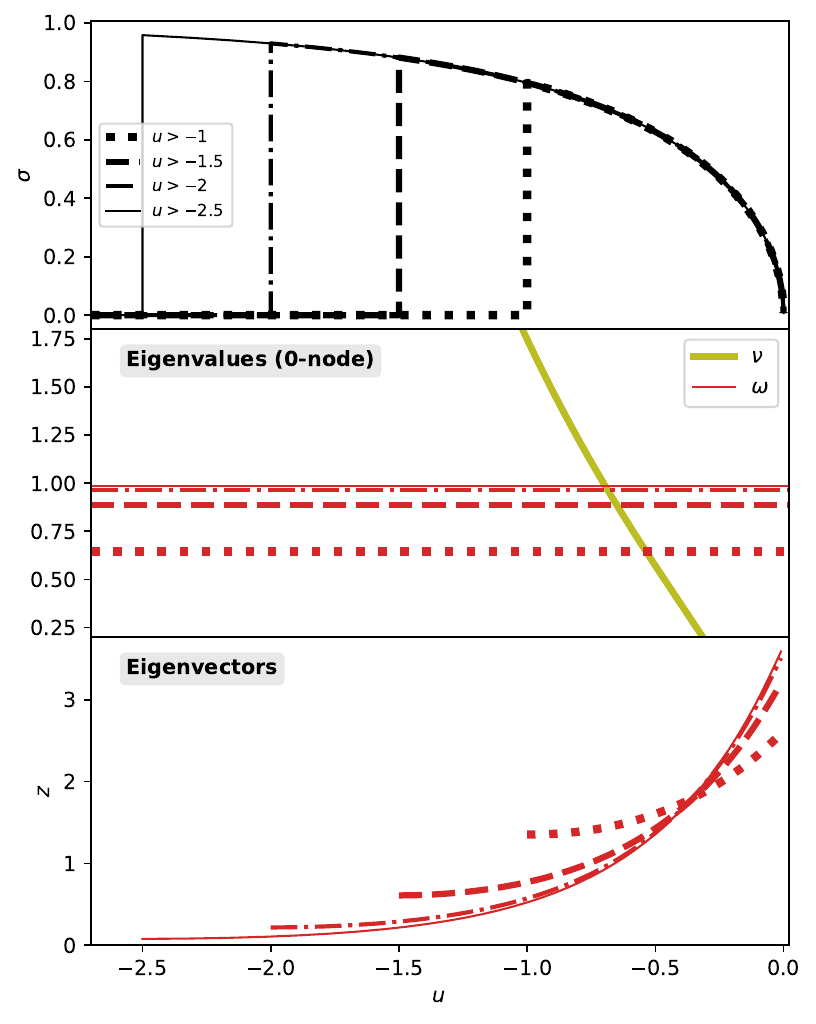}
\caption{{\bf Effect of Inner Edge:} 
The top panel shows  four $\sigma(u)$ profiles. They are the same
as the previous sharp edge case ($\gamma=0.5$), but here truncated at an  inner edge.  
The middle panel shows the effective potential $\nu$, and the eigenvalues $\omega$ of the 0-node
in each of the four profiles, similar to the top panel of Figure \ref{fig:drm}, but zoomed in. The bottom panel 
shows the four corresponding eigenvectors; unlike in Figure \ref{fig:drm}, the eigenvectors are not
scaled here by $\sigma$.}
\label{fig:u_cuts}
   \vspace{1.cm}
\end{figure}
The top panel of Figure \ref{fig:u_cuts} shows the four $\sigma$ profiles considered, which differ in the location of the
inner edge. In the bottom two panels, we show the resulting eigenvalues and eigenvectors for the zero-node mode.  
We see that once the inner edge is at $u_{\rm edge}\lesssim -1.5$, the eigensolution is independent of the  edge's
location, as previously claimed.  The reason for this independence is that the eccentricity mode is an 
edge mode. From the energy-level diagram (middle panel), it is trapped between 
the outer edge and
where the effective potential $\nu$ intersects 
$\omega$.\footnote{ Calling the mode an edge mode does not imply the $\sigma z$ profile in the evanescent zone is irrelevant. Each eigensolution is a global solution; the $\sigma z$ profile everywhere matters, but less for $\sigma z$ farther from the oscillatory zone. }

For the second case, we consider a narrow Gaussian ring,
\be
\sigma&=&  e^{-{u^2\over 2 s^2}}  \,. \label{eq:gauss}
\ee
 \cite{2001AJ....121.1776T}  finds that a Gaussian ring can support eccentric modes, provided
 it is sufficiently narrow ($s<0.07$). But that result, taken at face value, contradicts our finding that a mode can only be supported if the
 integral $\int (a/\sigma)du$ converges (Equation (\ref{eq:conv})). 
 For a Gaussian ring the integral
  diverges, whatever the value of $s$.
The apparent contradiction is resolved by noting that  the numerical experiments of
   \cite{2001AJ....121.1776T} restrict $|u|<5s$, in effect adding
   a sharp truncation
   at $|u|=5s$. To illustrate, we show in 
\begin{figure}
\centering
            \vspace{.5cm}
    \includegraphics[width=.5\textwidth]{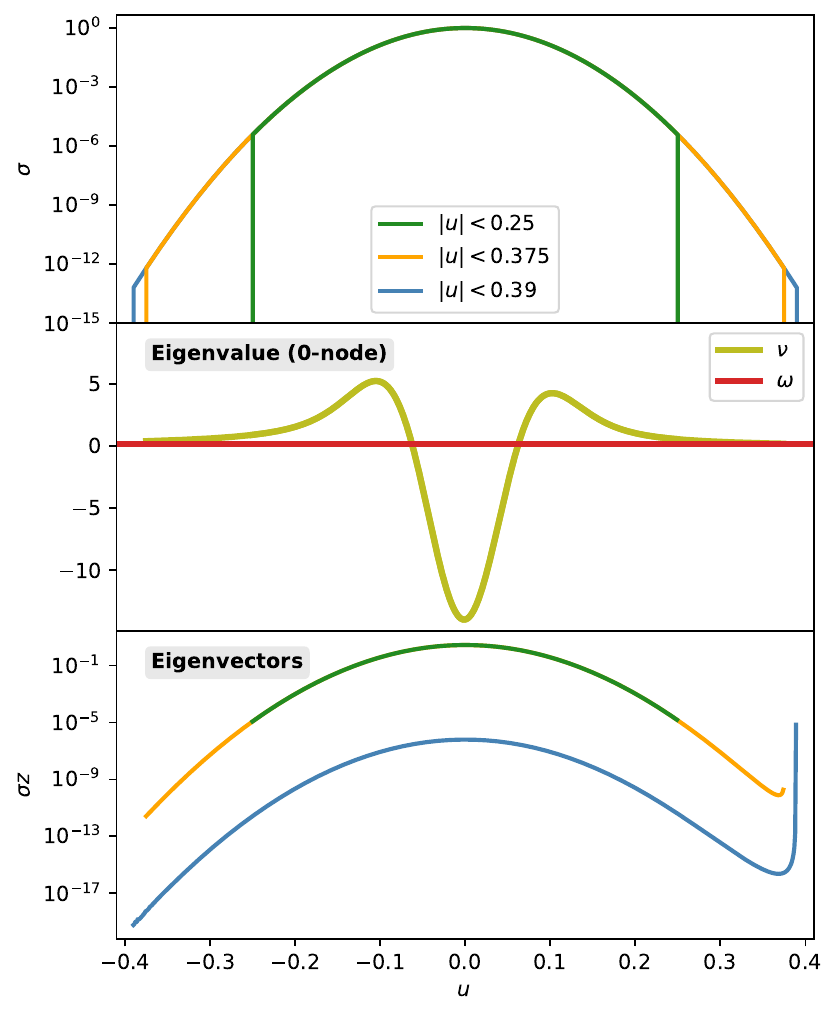}
\caption{{\bf Truncated Gaussian Ring:} 
The top panel shows  three Gaussian $\sigma(u)$ profiles with scale $s=0.05$ (Equation (\ref{eq:gauss})), and truncated at $|u|=0.25$, 0.375, and
0.39.
The first truncation  is equivalent to that in  \cite{2001AJ....121.1776T}.
The middle panel shows the energy-level diagram; the 
$\nu$ and $\omega$ for all three truncations are plotted, 
but the differences are too small to see.
The  bottom panel shows the zero-node eigenvectors for each of the three $\sigma$ profiles. 
This figure is made with resolution $\Delta=0.001$, but we have
confirmed that for smaller $\Delta$ there is no visible change. 
}
\label{fig:u_gauss}
   \vspace{1.cm}
\end{figure}
   Figure \ref{fig:u_gauss} the result from a $\sigma$ profile that
   is equivalent to the one in Tremaine, with $s=0.05$ and $|u|$ truncated
   at $5s$ (i.e., $|u|<0.25$). 
   For that case, Tremaine finds a well-behaved mode.
   The green curve in the top panel is the $\sigma$ profile, 
   and the one in the bottom panel is our resulting zero-node
   eigenvector, $\sigma z$.  We have confirmed that the shape of our $z$ is the same as his.
   We also show in the figure what happens when we increase the
   truncation  to $|u|=0.375$ and then $0.39$: the eigenvector shoots up dramatically near
   the outer boundary.
   Recalling that our eigenvectors are normalized to unit AMD, we
   see that at a truncation of 0.39, the AMD is dominated near the outer
   boundary. 
   When  the truncation  is increased beyond 0.39 by a very small
   amount,  the eigenvector  continues its extremely rapid rise, indicating
   that the eigenvector is pathological. 
   Hence an untruncated Gaussian ring cannot support linear modes.

\section{Discussion}

\label{sec:discuss}

\subsection{Neglected Effects}
\label{subsec:neglect}

We showed that self-gravitating disks  can only sustain an eccentricity 
if they have a sufficiently sharp edge; in particular, the edge must be sufficiently
sharp that $\int da/\sigma$ converges at the edge (Equation \ref{eq:conv}). 
To derive this result, we assumed 
that the disk is 2D, is forced
solely by gravity, and is governed by  linear secular theory. 
  Most or all of these assumptions must be relaxed
 in realistic disks.  Nonetheless,
 the dynamics discussed here should provide a 
 foundation for understanding more realistic disks. 
 Some of the most important additional
 effects are as follows:

\bi
\item Pressure: 
 For eccentric modes in a fluid disk, the strength of pressure relative to self-gravity is quantified by the
  dimensionless parameter 
  $h^2/\mu$, where 
  $h=(c/\Omega a)$, $c$ is the sound speed, and $\mu$ is the  local  ratio of disk mass to central body mass
   \citep{1999MNRAS.308..984L,2019ApJ...872..184L}. 
   Self-gravity dominates when
  \be
\mu \gtrsim h^2 \ \ \   {\rm (strong\ gravity)}. \label{eq:strong2}
  \ee
  In terms of Toomre's stability parameter, $Q\sim h/\mu$,
  that requirement reads
   \be
 Q\lesssim 1/h
 \label{eq:strong1}
  \ee
Disk stability 
adds an additional requirement,
$Q\gtrsim 1$ (i.e., $\mu\lesssim  h$). 
  Disks in the weak-gravity limit ($\mu\lesssim h^2$) 
   support purely pressure-dominated modes, 
  and the trapping of such modes is  much less sensitive to the disk's outer profile
      \citep[e.g.,][]{2019ApJ...882L..11L}.\footnote{
The difference between gravity and pressure dominated disks can be seen 
 from the functional form of the dispersion relation $\omega(k,u)$, graphically depicted as   contours in the dispersion relation map (DRM). Those contours 
  are the train tracks along which a mode travels.  For gravity-dominated modes the
 contours have a hyperbolic-like shape at the outer edge
 (Figure \ref{fig:drm}, bottom panel), and so a mode travelling outward  does not turn back until it reaches the edge. 
 Equivalently, reflection requires
zero group velocity, and since the 
  group velocity  is 
$\propto \Sigma$ at high $k$
(Equation (\ref{eq:kg1})), reflection requires
 $\Sigma = 0$ 
 (reached either continuously, as in the $\sigma$ profile of Equation (\ref{eq:profile}), or discontinuously, as in the step-function surface densities considered by \citealt{1999ASPC..160..307S})). 
 By contrast, for pressure-dominated modes the contours in the DRM are ellipses centered
 on the $k=0$ axis
 \citep[see Figure 5, top-left panel, in][]{2019ApJ...872..184L}. In following an ellipse, an outwardly travelling mode turns back
 before reaching the edge. 
 From the group velocity, 
  which is 
 $-h^2\Omega k$ in the pressure case, one
 may also infer that
 reflection occurs at $k=0$  
 \citep[the Lindblad resonance;][]{2019ApJ...872..184L}.
}
   \cite{2019ApJ...872..184L} study disks with pressure and self-gravity in both weak- and strong-gravity regimes, where the disks
   have soft edges. They find trapped pressure modes in the weak-gravity regime, and no modes (or more precisely, pathological very-small-wavelength modes)   in the
   very-strong-gravity regime, consistent with the results of the present paper and 
   \cite{2019ApJ...882L..11L}. 
   Were 
     \cite{2019ApJ...872..184L}
   extended to model sharp edges,  the disks would presumably
   support large wavelength  eccentric modes whether gravity was  in the weak or strong limit.  
   
\item Finite thickness, random velocities, and  softening: \black \cite{2003ApJ...595..531H} models the finite thickness of a disk with softening. 
Since softening acts somewhat like pressure \citep{2001AJ....121.1776T}, one might expect
that the same considerations discussed above for pressure would apply here, but with
$h$ now representing the disk's aspect ratio.  Non-fluid disks, such as stellar disks in galactic centers or debris disks, 
 could behave similarly, as they have scale-height $\sim v_{\rm random}/\Omega$.  Softening can also be used to treat cases where orbits cross, e.g., to obtain precession rates of orbit-crossing test particles embedded in disks.  

\item 
External potential:   It would be of interest to extend the study in the present paper to include an external potential, 
which might help trap an eccentric mode, even if the disk has a soft edge.
In protoplanetary
disks or planetary rings, 
 the 
external potential would typically come from 
a companion 
(planet or moon), or a bulge on the central body. 
\cite{2001AJ....121.1776T} includes externally-driven precession, 
which indeed helps traps his modes.  
But for simplicity he does not model a companion's potential.
Instead, he 
scales the $K_1$ term in our master equation (Eq. \ref{eq:eom})
by a constant factor.

Tremaine finds that some of his modes have negative eigenfrequencies, whereas Laplace-Lagrange (L-L) theory should always produce positive eigenfrequencies for eccentric modes.\footnote{For an inclined (warped) disk, L-L eigenfrequencies are negative.}
He argues on this basis that L-L
does not give the correct answer for a continuous disk.
We contend that L-L is correct 
(see the end of our section \ref{sec:master}), 
and 
that it is his 
external precession
 that is the source of
the negative eigenfrequencies.
To be specific, 
his external precession 
is equivalent to that from a ``frozen''
(time-independent) potential
of an axisymmetric disk
having a scaled version of
our disk's surface density profile.
Thus his setup requires
orbits in the eccentric 
disk to 
  cross those in the frozen
  circular 
  disk. It is that orbit-crossing that makes
L-L invalid, and the eigenfrequencies
negative.

\item Nonlinearity: In disks with a soft edge, an eccentric disturbance becomes nonlinear
as it leaks into the low density region (Figure \ref{fig:tevol}, right panel).  In order
to find out what happens next, one must include nonlinear effects. 
\cite{1999MNRAS.308..984L} study the nonlinear dynamics of forced eccentric modes
in power-law disks.  

\item Excitation and Decay: 
Even if a disk can sustain eccentric modes, that does not guarantee that it is eccentric. 
There must additionally be a mechanism to excite the eccentricity. 
One possibility for excitation is a 
companion such as a planet or binary star. 
A second is instability.
The indirect potential can lead to instability for sufficiently massive disks, even
if the disk is Toomre-stable
\citep{1989ApJ...347..959A,1990ApJ...358..495S}. Instability can also be triggered by a locally isothermal equation of state
\citep{2015MNRAS.448.3806L},
 by a finite cooling time  \citep{2021ApJ...910...79L}, or by 
 non-axisymmetric irradiation (``shadows'') 
 \citep{2024ApJ...976....5Q}.

After an eccentric mode is excited, it can be damped by viscosity.  
Pressure-dominated modes are relatively immune to viscous damping: their decay time is comparable
to the viscous time of the disk \citep{2016MNRAS.458.3221T,2019ApJ...882L..11L}, 
which should be comparable to the disk's lifetime. The viscous decay time for self-gravity-dominated
modes is likely also long.
 
 \item Warped disks with self-gravity:  the governing equation for a warped
 disk is nearly identical to our master equation  (Equation \ref{eq:eom}).
 The only differences are that $z$ in that equation should represent the
 complex inclination ($z\rightarrow ie^{i\Omega}$ in standard notation) 
 rather than the complex eccentricity, and
 one must  replace $K_2\rightarrow K_1$ \citep{1999ssd..book.....M}.
  Hence our dispersion relation  
 and analysis carries through virtually unchanged.
  \cite{2018MNRAS.475.5070B} investigates warped disks with self-gravity by adopting
  a number of simplifications. 
He considers a profile with $\sigma/a=$ const, and approximates
 self-gravity by considering only nearest-neighbor interactions. His resulting disperion relation is
 $\omega=$ const$\times k^2$ (his Equation 33), which is the same as for
 the potential-free Schr\"odinger equation. 
 Contrasting his dispersion relation with ours
 (Equation \ref{eq:dr}), we see that his corresponds to dropping the $\nu$ term, and taking the small
 $k$ limit of  $\K(k)$. Such an approximation may be justifiable under specific conditions. 

\ei

\subsection{Potential Applications}
\label{sec:apps}

\bi
\item Protoplanetary Disks: 
Some protoplanetary disks have eccentric dust cavities, including
MWC 758 and  AB Aur \citep{2018ApJ...860..124D,2021AJ....161...33V}, 
and 
others \citep{2024AJ....167..115J}. 
A standard explanation
for the origin of the eccentricity is a massive planet, which can excite the
disk's eccentricity when the planet is more than a few Jupiter masses 
\citep{2001A&A...366..263P,2006A&A...447..369K,2021ApJ...918L..36D}, 
although other explanations have also been proposed \citep[e.g.,][]{2024ApJ...976....5Q}.
Some young (Class 0) disks exhibit asymmetry 
\citep{2023ApJ...958...60T,2023ApJ...951...10V},  which has been attributed to disk
eccentricity 
instigated by non-axisymmetric accretion during cloud collapse
\citep{2024A&A...689L...9C}. 

Many protoplanetary disks are sufficiently massive that self-gravity dominates
pressure for the dynamics of the eccentric modes (Equation \ref{eq:strong1}).
We would expect that such disks could only be eccentric if their edges were extremely sharp.
However, multiple effects not considered here could alter that conclusion, including
nonlinearity  and the relative of importance of pressure and self-gravity
near the edge
 (Section \ref{subsec:neglect}). Those effects 
must be addressed before making definite predictions.

\item Planetary Rings: Uranus's $\epsilon$ ring is eccentric, 
despite
differential precession from Uranus's quadrupole
which would quickly destroy the eccentricity in the absence of a counterbalancing force.  
\cite{1979AJ.....84.1638G} proposed
that the eccentricity can be  maintained by self-gravity.  Other rings (e.g., the Uranian $\alpha$ and $\beta$ rings, and Saturn's Maxwell ring) have broadly similar properties. 
The $\epsilon$ ring is observed to have sharp edges, which are maintained by the
 shepherd moons  Cordelia and Ophelia \citep{1979Natur.277...97G,1982Natur.299..209B,1989Icar...80..344B, 1987AJ.....93..724P, 1987AJ.....93..730G}, and it and other rings 
 lie in the strong gravity regime by a  large margin ($Q\sim 1$ and $h\ll 10^{-6}$; see Equation \ref{eq:strong1}). 
Plausibly, sharp edges are essential for these rings to maintain their eccentricities. 
However,  some of the effects neglected in this paper   play  important roles in narrow eccentric rings, including
nonlinearity and pressure. 
Nonlinearity is included in the model of 
 \cite{1979AJ.....84.1638G}, and the same nonlinear equations are modified by pressure forces in the model 
by \cite{2000ApJ...540.1084C}.  The latter authors showed that pressure forcing, which is especially strong near ring edges, 
is needed to quantitatively account for the  
measured eccentricities and masses of the 
Uranian rings. 
See also \citet{2003ApJ...599..675C} who extended these considerations to treat inclinations.

\item Stellar Disks in Galactic Nuclei:  M31 has a double nucleus that can be explained as an eccentric disk of stars 
\citep{1995AJ....110..628T}.  A variety of other galaxies also host double nuclei  \citep{1996ApJ...471L..79L,2000A&A...364L..47T,2006ApJ...651L..97D}.
As discussed above, \cite{1999ASPC..160..307S} modelled eccentric nuclear disks with Laplace-Lagrange theory. 
They adopted power-law surface density profiles with infinitely sharp edges, and they successfully found
eccentric modes. That result is understandable 
from our theory.   
Observationally, M31 has an aspect ratio $h\sim 0.4$, and 
a ratio of disk to black hole mass of $\mu\sim 0.15$
\citep{2003ApJ...599..237P}. 
Thus from Equation (\ref{eq:strong2}),  the effect of finite scale height may be important.

\ei

\section{Summary}
\label{sec:sum}

 A Keplerian disk in the strong-gravity regime ($\mu\gtrsim h^2$) must have a sharply truncated  edge in order to be
eccentric.  The surface density $\Sigma$ must hit zero at the edge,
and the integral in Equation (\ref{eq:conv}) must converge, which means that the
time 
 for a wavepacket to reach the edge must be finite.  
 We may approximate that requirement as
\be
 \int^{a_{\rm edge}}{da\over \Sigma(a)} < \infty \ . \label{eq:convint}
\ee
Asymptotic outer edges, e.g., of the form $\Sigma\sim e^{-a^\xi}$ for any $\xi$, 
are insufficiently sharp to support an eccentric mode.
\cite{1969ApJ...155..747H} similarly show that the integral in Equation (\ref{eq:convint}) must
converge in order for 
a galactic disk to support a warp.  They draw an   analogy with waves on a lake, 
which last much longer when the lake is bounded by vertical cliffs than by beaches. That is because waves are reflected by cliffs but not beaches. 
For  a disk with a sufficiently sharp  edge, the wave is reflected by the edge, whereas
if the edge is soft
 the eccentric disturbance leaks out into the edge region, 
and its energy (or really, AMD) is  lost.

For our  fiducial sharp profile (Equation (\ref{eq:profile}) with $\gamma=0.5$), the normal modes   
are edge modes. Their AMD is dominated near the outer boundary, and
the trapping zone is within less than a decade of the  edge (Figure \ref{fig:drm}, top panel). 
For different surface density profiles, the modes' AMD can be dominated at the inner edge \citep{1999ASPC..160..307S}, in which case Equation (\ref{eq:convint}) should apply at the inner edge.

We also showed that two-dimensional disks supported by self-gravity may be analyzed
without introducing artificial  softening.  
Nonetheless, 
softening may still be useful as a model for additional effects such as pressure or a finite vertical
extent
  \citep{2001AJ....121.1776T,2003ApJ...595..531H}. 
In the appendix, we  derived a  discretized equation for the eccentricity
that does not use softening, 
conserves
a discretized AMD, and
  converges rapidly with grid spacing  (Equation (\ref{eq:zdotdisc})). 
Some previous studies have introduced  softening-free algorithms \citep{1996ApJ...460..855L,2019ApJ...872..184L}.  But the algorithm in Equation
(\ref{eq:zdotdisc}) is more straightforward to implement, and we have  verified 
that its accuracy is $O(\Delta^2)$ (Figure \ref{fig:Idisc}).

 Finally, we  derived a new and simple dispersion relation for eccentric disturbances
(Equation (\ref{eq:dr})).
Unlike WKB, it is valid at any wavenumber.
It 
 allows one
 to show that the integral of Equation (\ref{eq:convint}) controls the travel
 time to the edge. More generally, it allows one
 to plot energy-level diagrams that are familiar from solving
 Schr\"odinger's equation. The diagrams provide a graphical depiction of where
eccentric modes are trapped, and of  their precession frequencies.

\black

\section*{\bf Acknowledgements}
We thank Scott Tremaine for 
insightful and detailed feedback on this work 
and for telling us about
\cite{1969ApJ...155..747H}.
We also thank
an anonymous referee for a constructive report
and 
 Antranik Sefilian for helpful pointers to the literature on softening and eccentric modes.
Y.L. acknowledges NASA grant 80NSSC23K1262.
E.C. acknowledges a Simons Investigator grant, a Berkeley Discover grant for undergraduate education, and a Berkeley Letters \& Science Summer Undergraduate Research Fellowship (SURF) awarded to Z.Y.

\appendix
\section{\bf Discretization}
\label{sec:disc}

We write the master equation  (Equation (\ref{eq:eom}))  as 
\be
\dot{z} = i{1\over a}\I(u)  \ ,
\ee
 where
\be
\I(u)&\equiv & \int\left(K_1z-K_2z'  \right) {\sigma' }du' 
\label{eq:idef}
\ee
and discretize $\I$ on a grid in $u$. The spacing in $u$ is taken to be uniform, and denoted $\Delta$. 
For the purposes of the derivation, we
 imagine that  $z$ and $\sigma$ 
are smooth functions of $u$, and that their values on the  grid are known.
Gridpoints are labelled $u_j$, for integer $j$, and the values of $z$ and $\sigma$ on the grid are
$z_j$ and $\sigma_j$.

As explained in the body of the paper, the $K_m\equiv K_m(u-u')$ diverge as $u'$ approaches $u$. 
But $\I$ remains finite, despite that divergence, because as $u'\rightarrow u$, 
then both $z'\approx z$ and $K_2\approx K_1$. To separate out those 
two effects, we first decompose
\be
\I&=&
{\cal A}+{\cal B}
 \\
{\cal A}
&=& \int K_2(z-z') {\sigma' } du'
\\
{\cal B}&=&
z \int \left(K_1-K_2\right) {\sigma' } du' 
\label{eq:ib}
\ee
and discretize each of the components $\A$ and $\B$ in turn. 

The naive discretization of $\A_j\equiv \A(u_j)$ is, in obvious notation,
\be
\A_j^{\rm naive} =  \Delta \sum_{k\ne j} K_{2,jk}(z_j-z_k) \sigma_k \label{eq:naive}
\ee
But one must be careful at the gridpoint $k=j$, since $K_2$ diverges there, while
 $z_j-z_k$  vanishes. 
 In the naive method in Equation (\ref{eq:naive}), we
simply discarded the $k=j$ point. 
To examine whether that is sufficiently accurate,  we write the contribution to $\A(u)$ within a distance $\Delta/2$ as
\be
\A^{\Delta}(u) 
&=&\int_{-\Delta/2}^{\Delta/2} K_2(v)(z-z_+)   \sigma_+  dv \\
&=& \int_0^{\Delta/2} K_2(v)\left[\left(z-z_+  \right)\sigma_++(z-z_-)\sigma_-\right]dv
\label{eq:iadel}
\ee
where in the first line we changed the integration variable to $v\equiv u'-u$, and defined $z_+=z(u+v)$, and
similarly  $\sigma_+=\sigma(u+v)$; and in the second line we combined positive
and negative values of $v$, taking advantage of the symmetry of $K_2$, and defined
$z_-=z(u-v)$, and similarly for $\sigma_-$.
As we are concerned presently with small $\Delta$, we Taylor-expand the integrand in $v$, which gives
\be
\A^{\Delta}(u) 
 &\approx&
\left[ z {d^2\over du^2}\sigma-{d^2\over du^2}(z\sigma )\right] \int_0^{\Delta/2} K_2(v)v^2 dv \ .
\ee
Since  $K_2(v)\propto 1/v^2$ at small $v$, the  integral in that expression converges. But the integrand
is flat in $v$, and thus discarding the $k=j$ contribution to Equation (\ref{eq:naive}) leads to a fractional error of
$\sim \Delta$. 
That is indeed small; but one can do better.
One can correct the leading error in a nearly trivial way by changing the discretization
from Equation (\ref{eq:naive}) to
\be
\A_j =  \Delta \sum_k K_{2,jk}(z_j-z_k) \sigma_k F_{jk} \ , 
\ee
where
\be
F_{jk} =
\begin{cases}
0   & \text{if } k = j, \\
1.5 & \text{if }  k=j\pm 1, \\
1   & \text{otherwise.}
\end{cases}
\label{eq:fjk}
\ee
The factor of 1.5  compensates for the discarded $k=j$ term, taking advantage
of the fact that the integrand is nearly independent of $v$.
The error in the discretization is now $\sim \Delta^2$, as confirmed experimentally in the body of the paper. 

Next, we turn to $\B$ (Equation (\ref{eq:ib})), which
we discretize  as follows
\be
\B_j = z_j  \Delta \sum_k L_{jk}\sigma_k 
\ee
where
\be
L_{jk}
&=&{1\over\Delta}\int_{u_k-u_j-\Delta/2}^{u_k-u_j+\Delta/2} (K_1(v)-K_2(v)) dv \ .
\label{eq:gjk}
\ee
These integrals are precomputed numerically.
For the diagonal component ($L_{jj}$), the integrand diverges at $v\sim 0$. But  the divergence is sufficiently weak ($\sim \ln |v|$)
that the integral converges. 
If the functional form of $\sigma(a)$ is known, 
one could alternatively integrate Equation (\ref{eq:ib}) exactly, i.e., without pulling the $\sigma'$ out of the integral. But 
the increased accuracy is of little benefit,  as we have already discretized $\sigma$ for $\A$.

Collecting results,
we have shown that the integral $\I$, as defined in Equation (\ref{eq:idef}), is
 to be discretized as
\be
\I_{\rm discrete}(u_j)&=& \Delta \sum_k  K_{2,jk}F_{jk}(z_j-z_k)\sigma_k + z_j L_{jk}\sigma_k
\label{eq:idisc}
\ee
where $F_{jk}$ and $L_{jk}$ are provided by Equations (\ref{eq:fjk}) and (\ref{eq:gjk}).
The discretized equation of motion is then
\be
\dot{z}_j = i{1\over a_j}
\Delta \sum_k  K_{2,jk}F_{jk}(z_j-z_k)\sigma_k + z_j L_{jk}\sigma_k
\label{eq:zdotdisc}
\ee

\subsection{\bf Matrix Formulation and Eigenvalue Equation}

In order to solve  Equation  (\ref{eq:zdotdisc}) numerically, 
we   convert it to  matrix form.
We  represent $z_j$ as a column vector $\bld{z}$, and similarly for $a_j$ and $\sigma_j$:
\be
z_j&\leftrightarrow& \bld{z} \\
a_j &\leftrightarrow& \bld{a} \\
 \sigma_j &\leftrightarrow& \bld{\sigma}  \ .
\ee
 And we represent $K_{2,jk}F_{jk}$ and $L_{jk}$ as matrices:
\be
 K_{2,jk}F_{jk}&\leftrightarrow& \bld{K_2}
 \\
  L_{jk}&\leftrightarrow& \bld{L} 
\ee
The matrix form of Equation (\ref{eq:zdotdisc}) is then
\be
 \bld{\Lambda_a \dot{z}}
 = i\Delta  
 \left( 
 \bld{\Lambda_{K_2\sigma+L\sigma}}-\bld{K_2 \Lambda_\sigma}
 \right) \bld{z} \ ,
\ee
where we
 represent a diagonal matrix that has a vector (e.g., $\bld{a}$) on its diagonals as
\be
\bld{\Lambda_a}\equiv& {\rm DIAG}({\bld a}) 
\ee
We convert this to a more  standard form by 
multiplying both sides by $\bld{\Lambda_\sigma}$, 
\be
\bld{\Lambda_\sigma\Lambda_a \dot{z}}
 = i\Delta  
 \left( 
 \bld{\Lambda_\sigma\Lambda_{K_2\sigma+L\sigma}}-\bld{\Lambda_\sigma K_2 \Lambda_\sigma}
 \right) \bld{z} \ .
  \label{eq:eomdisc}
\ee
Now the
 matrix multiplying $\bld{{z}}$ is real and symmetric, as is the matrix
multiplying $\bld{\dot{z}}$. 
One consequence of these symmetries is that
the discretized equation can be seen by inspection to exactly conserve a discretized AMD 
\be
{\rm AMD}_{{\rm discrete}}\equiv \Delta \bld{z^\dagger}\bld{\Lambda_\sigma\Lambda_a}\bld{z} = \Delta\sum_k |z_k|^2\sigma_ka_k
\ee
(compare with Equation (\ref{eq:amd})).

To obtain the eigensolutions of Equation (\ref{eq:eomdisc}), 
we set 
\be
\bld{z} = \bld{\hat{z}} e^{i\omega t}  \ ,
\ee
which gives the eigenvalue equation 
\be
\omega \bld{\Lambda_\sigma\Lambda_a \hat{z}}
 = \Delta  
 \left( 
 \bld{\Lambda_\sigma\Lambda_{K_2\sigma +L\sigma}}-\bld{\Lambda_\sigma K_2 \Lambda_\sigma}
 \right) \bld{\hat{z}} \ . \label{eq:eommatrix}
\ee
Because of the aforementioned symmetries of the matrices on the two sides of the equation, 
 the eigenvalues and eigenvectors are real, and the eigenvectors are orthogonal when they sandwich the matrix $\bld{\Lambda_\sigma\Lambda_a}$ 
 \citep[e.g.,][]{goldstein1980classical}.
  To be explicit, if we label different eigenvectors with
 subscript $l$ and $m$, then
 \be
 \Delta \bld{\hat{z}^\dagger_l}\bld{\Lambda_\sigma\Lambda_a} \bld{\hat{z}_m} = \delta_{lm}
 \label{eq:ortho}
 \ee
Our insertion of $\Delta$ in Equation (\ref{eq:ortho}) is arbitrary. 
But it is convenient because it sets the normalization convention to be that
the discretized AMD in a mode is unity; i.e., 
\be
{\rm AMD}_{\rm discrete}\equiv \Delta \sum  |z_k|^2\sigma_ka_k = 1 \label{eq:amdnorm}
\ee
where $z_k$ are the values of $z$ on the grid, for any mode.

\bibliography{edgebib}

\bibliographystyle{apj}

\end{document}